\documentclass[aps,prc,letterpaper,twoside,tightenlines,nofootinbib,showpacs
,preprint,superscriptaddress]{revtex4-1}
\usepackage{setspace}
\usepackage{epsfig,float}
\usepackage{amsmath}
\begin{document}

\newcommand{\Ima}{\textrm{Im}}
\newcommand{\Rea}{\textrm{Re}}
\newcommand{\mev}{\textrm{ MeV}}
\newcommand{\gev}{\textrm{ GeV}}
\newcommand{\rb}[1]{\raisebox{2.2ex}[0pt]{#1}}

\title{Baryon states with open charm in the extended local hidden gauge approach}

\author{W. H. Liang}
\email{liangwh@gxnu.edu.cn}
\affiliation{Department of Physics, Guangxi Normal University, Guilin, 541004, People's Republic of China}

\author{T. Uchino}
\author{C. W. Xiao}
\author{E. Oset}
\affiliation{
Departamento de F\'{\i}sica Te\'orica and IFIC, Centro Mixto Universidad \\de Valencia-CSIC, Institutos de Investigaci\'on de Paterna, Apartado 22085, 46071 Valencia, Spain}

\date{\today}

\begin{abstract}

In this paper we examine the interaction of $D N$ and $D^* N$ states, together with their coupled channels, by using an extension of the local hidden gauge formalism from the light meson sector, which is based on heavy quark spin symmetry. The scheme is based on the use of the impulse approximation at the quark level, with the heavy quarks acting as spectators, which occurs for the dominant terms where there is the exchange of a light meson. The pion exchange and the Weinberg-Tomozawa interactions are generalized and with this dynamics we look for states generated from the interaction, with a unitary coupled channels approach that mixes the pseudoscalar-baryon and vector-baryon states. We find two states with nearly zero width which are associated to the $\Lambda_c(2595)$ and $\Lambda_c(2625)$. The lower state, with $J^P = 1/2^-$, couples to $D N$ and $D^* N$, and the second one, with $J^P = 3/2^-$, to $D^* N$.  In addition to these two $\Lambda_c$ states, we find four more states with $I=0$, one of them nearly degenerate in two states of $J=1/2,\ 3/2$. Furthermore we find three states in $I=1$, two of them degenerate in $J=1/2, 3/2$.

\end{abstract}


\maketitle

\section{Introduction}

In dealing with hadronic states involving heavy quarks (charm or beauty) the heavy quark spin symmetry \cite{isgur,neubert,manohar,wise} plays an important role and serves as a guiding principle to proceed with calculations. Heavy quark spin symmetry (HQSS) has been applied to calculate baryon spectra in Refs. \cite{GarciaRecio:2008dp,Flynn:2011gf,GarciaRecio:2012db,Romanets:2012hm,Guo:2013xga,xiaojuan,
carmen,olena}. The basic idea behind these works is to use HQSS to reduce the freedom in the interaction, which is then written in terms of a few parameters which are adjusted to some experimental data. Then predictions on spectra of baryons with charm or beauty, or hidden charm and beauty are made. In Ref. \cite{GarciaRecio:2008dp} an SU(8) spin-flavor scheme is used, to account for the spin symmetry, in order to obtain the interaction, and a coupled channels unitary approach is implemented to obtain poles in the scattering matrices, which correspond to the baryon resonance states. In particular the $\Lambda_c(2595)$ state is obtained and shown to couple largely to the $D^* N$ channel. In  Ref. \cite{Romanets:2012hm} the SU(8) scheme is once again used, but with some symmetry breaking, to match with an extension of the Weinberg-Tomozawa interaction in SU(3). Among other resonances, the states $\Lambda_c(2595)$ ($J^P=1/2^-$) and $\Lambda_c(2625)$ ($J^P=3/2^-$) are obtained.

  Further steps on the relationship of the Weinberg-Tomozawa interaction and HQSS are given in Refs.
\cite{xiaojuan,xiaooset}, where it is shown that this interaction, which stems from the exchange of vector mesons in the local hidden gauge approach \cite{hidden1,hidden2,hidden4} (see also Ref. \cite{hideko} for practical rules), fulfills the HQSS. Indeed, the dominant terms correspond to the exchange of light vectors (the exchange of heavy vectors is suppressed), the heavy quarks act as spectators and hence the interaction does not depend on them. HQSS is then automatically fulfilled. Another step forward in this direction was given in Ref. \cite{liang}, where using the impulse approximation at the quark level, the Weinberg-Tomozawa interaction of the SU(3) sector was extended to the heavy bottom sector and it was shown that the result is equivalent to the plain use of the local hidden gauge approach extended to SU(4)\footnote{Actually one only needs to use SU(3) once the heavy quarks act as spectators.}. It was also shown in Ref. \cite{liang} that the case of pion exchange requires a renormalization due to the field normalization of the mesons, which, however, is automatically implemented in the Weinberg-Tomozawa term. Another novelty in Ref. \cite{liang} was to realize that a better measure of the relevance of different channels in the wave function of the states is to look at the wave function at the origin, rather than the coupling from the residues at the pole of the different amplitudes. These findings are most welcome and a posteriori justify the approaches used in many works where the Weinberg-Tomozawa interaction has been used in the heavy quark sector
\cite{Tolos:2005ft,Lutz:2005vx,angelsmizu,Tolos:2007vh,Wu:2010jy,Wu:2010vk,wuzou}, or variants of it, as in the J\"ulich model \cite{Haidenbauer:2007jq,Haidenbauer:2010ch}.

   In Ref. \cite{liang} the states  $\Lambda_b(5912)$ and $\Lambda_b(5920)$ were obtained, among others, using a unitary scheme with coupled channels and the dynamics based on the local hidden gauge approach. These states, in $J^P=1/2^-, 3/2^-$ respectively, were naturally obtained in the $\bar B^* N$ and coupled channels sector, and the difference of masses comes from the different weight of the intermediate $\bar B N$ states, which are accounted for by means of box diagrams mediated by pion exchange.

In the charm sector we have states which bear some similarity with those states in the beauty sector. They are the $\Lambda_c(2595)$ ($J^P=1/2^-$) and $\Lambda_c(2625)$ ($J^P=3/2^-$) states which have been observed in various experiments \cite{pdg}. It is tempting to see if a similar explanation can be given in this case, or see if those states call for a different explanation. We anticipate that although there are similarities, there are also differences and, while the $\Lambda_c(2625)$ ($J^P=3/2^-$) state is mostly tied to the $D^* N$ channel, the $\Lambda_c(2595)$ state has an important coupling to both the $DN$ and $D^* N$ channels. The existence of these two states allows us to put constraints in the free parameters of our model and after this we make predictions for other states in isospin $I=0$ and $I=1$.

\section{Formalism}
\subsection{The construction of the interaction}

\begin{figure}[tb]
\epsfig{file=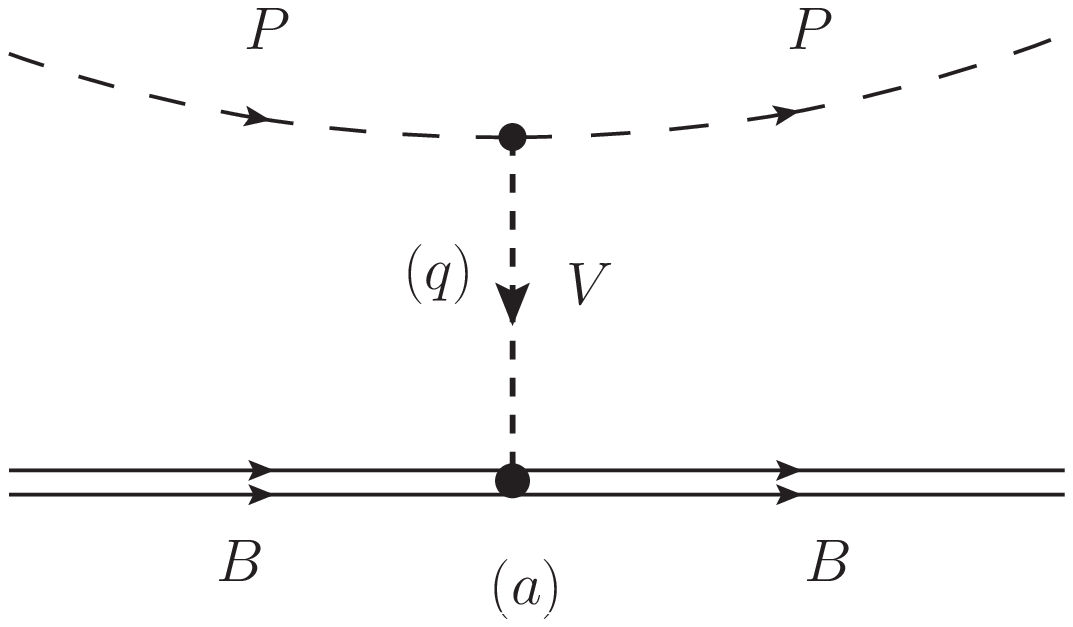, width=5.5cm} ~~~~~\epsfig{file=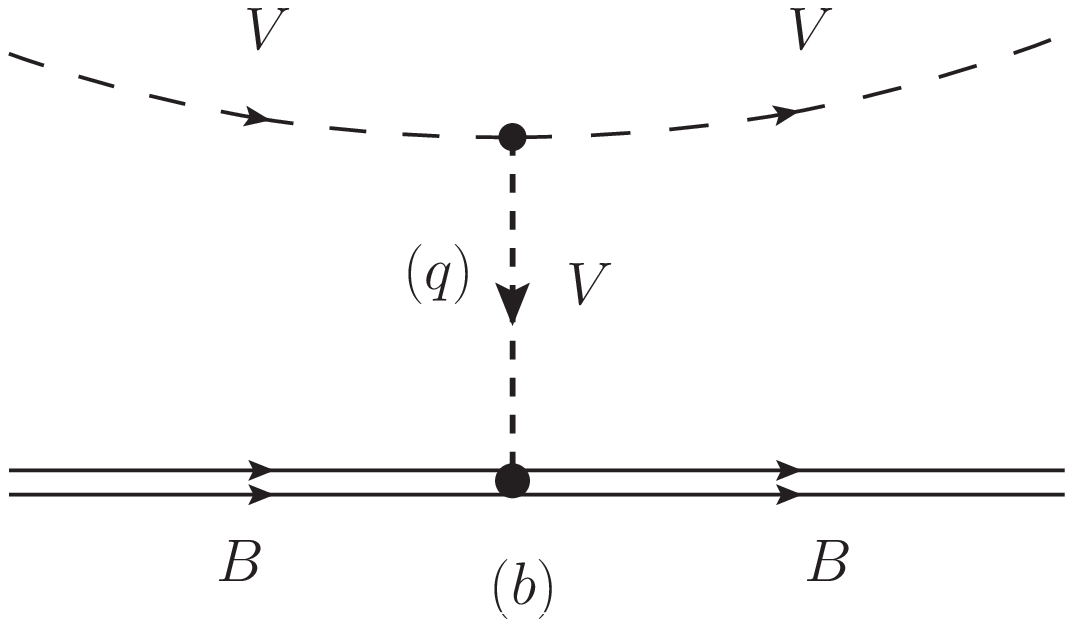, width=5.5cm}
\caption{Diagrammatic representation of the pseudoscalar-baryon interaction (a) and vector-baryon interaction (b).}\label{fig:f1}
\end{figure}

We take as basis of states $\pi \Sigma_c$, $\pi \Lambda_c$, $\eta \Lambda_c$, $\eta \Sigma_c$ and $D N$ which can couple to $I=0, ~1$. Similarly, we consider $D^* N$, $\rho \Sigma_c$, $\omega \Lambda_c$ and $\phi \Lambda_c$. As a singular case, we also consider the single channel $\pi \Sigma_c^*$, with $\Sigma_c^* = \Sigma_c^* (2520)$, belonging to a decuplet of $3/2^+$ states. In the local hidden gauge approach in SU(3) \cite{hidden1,hidden2,hidden4} the meson-baryon interaction proceeds via the exchange of vector mesons as depicted in Fig. \ref{fig:f1}.
In Ref. \cite{liang}, this approach was generalized to SU(4). As discussed in Ref. \cite{liang}, when we exchange a light vector meson in Figs. \ref{fig:f1}(a) and \ref{fig:f1}(b), the heavy quarks of the meson or the baryon are spectators and, as a consequence, the interaction does not depend on their spin-flavor. Technically, the interaction of the diagrams of Fig. \ref{fig:f1} can be obtained using SU(3) symmetry considering $u, ~d, ~c$ quarks, since we do not consider states with strangeness or hidden strangeness. Hence, all the matrix elements of the interaction are identical (except for the mass or energy dependence) to those of the interaction of the analogous states $\pi \Sigma$, $\pi \Lambda$, $\eta \Lambda$, $\eta \Sigma$, $\bar{K} N$, $\bar{K}^* N$, $\rho \Sigma$, $\omega \Lambda$, $\phi \Lambda$ and $\pi \Sigma^*$, which have been studied in Refs. \cite{angels,sarkar}.

The transition potential from channel $i$ to channel $j$ is given by \cite{bennhold}
\begin{equation}
V_{ij} = -C_{ij} \frac{1}{4f^2} (2 \sqrt{s} - M_{B_i} - M_{B_j}) \sqrt{\frac{M_{B_i} + E_i}{2 M_{B_i}}} \sqrt{\frac{M_{B_j} + E_j}{2 M_{B_j}}},
\label{eq:vij}
\end{equation}
with $f$ the pion decay constant, $M_{B_i}, ~E_i$ ($M_{B_j}, ~E_j$) the mass, energy of baryon of $i$ ($j$) channel. We take $f=f_\pi = 93\mev$ since we exchange light vector mesons. The $C_{ij}$ coefficients are evaluated in Refs. \cite{angels,sarkar} and we quote them in the Appendix. We note that, according to Ref. \cite{liang}, one must use $f_{\pi}$, since the corrections due the consideration of heavy hadrons are automatically taken into account in the energy dependence of the interaction (note that Eq. (\ref{eq:vij}) provides a relativistic version of the sum of the two external meson energies).

We should note that in the light sector, the local hidden gauge approach, with the exchange of vector mesons, neglecting the momentum transfer, leads to the chiral Lagrangians. When we exchange light vectors in the heavy quark sector, the heavy quark acts as spectator, and we can make a mapping of the results from the light sector without having to use SU(4) symmetry. For transitions that involve the exchange of $D^*$, we use elements of SU(4) as pointed above. Yet, those terms play a minor role since the transitions are suppressed due to the heavy mass of the vector exchanged. We are aware that SU(4) can be badly broken, but previous studies on the source of the breaking have concluded that SU(4) symmetry is actually quite good in the elementary vertices. In this sense it is interesting to recall that the radiative decays of $D, D_s, J/\psi$ are well described using the SU(4) couplings (see chapter 6 of Ref. \cite{gamermannthesis}). The large breaking in many processes is associated to the effect of the different masses in the contribution of the diagrams associated to them (see section IID of Ref. \cite{Wu:2010vk}).

We shall also use $\pi$ exchange in the transition from pseudoscalar-baryon ($PB$) to vector-baryon ($VB$) states, which is also a part of the hidden gauge formalism. Yet, one may wonder why not to consider $\sigma$ exchange which is relevant in nuclear physics. In  chiral dynamics the $\sigma$ [$f_0(500)$] is a state that appears from the interaction of pions \cite{npa, pelaez, kaiser}. From this perspective $\sigma$-exchange  was evaluated in Ref. \cite{mizobe} through the exchange of two interacting pions. This picture has also been recently used in the heavy quark sector to see the effect of $\sigma$-exchange in the interaction of heavy hadrons \cite{jorgifran,dos}, with the result that it is small compared with the exchange of $J/\psi$ (heavily suppressed) and, a fortiori, with respect to the exchange of light vectors. The different strength with respect to the $\sigma$ exchange in $NN$ interactions must be looked in the triangular vertices coupling the $\pi \pi$ system to a nucleon or to a meson, which involve the $\pi NN$ Yukawa coupling and the $VP \pi$ coupling respectively.

For pseudoscalar mesons and $J^P=$ $1/2^+$ baryons we have the coupled channels $D N$, $\pi \Sigma_c$, $\eta \Lambda_c$ in $I=0$ and the $C_{ij}$ coefficients are given in Table \ref{tab:vij0} of the Appendix.

In $I=1$ we have the channels $D N$, $\pi \Sigma_c$, $\pi \Lambda_c$, $\eta \Sigma_c$ and the $C_{ij}$ coefficients are given  in Table \ref{tab:vij1} of the Appendix.
We can see that the interaction in $I=0$ is stronger than that in $I=1$ and one has more chances to bind states in $I=0$.

Similarly for $VB$ with $I=0$ we have the channels $D^* N, \rho \Sigma_c, \omega \Lambda_c, \phi \Lambda_c$ and the $C_{ij}$ coefficients are given in table \ref{tab:vij0j13}  of the Appendix. For $VB$ with $I=1$, the channels are $D^* N, \rho \Sigma_c, \rho \Lambda_c, \omega \Sigma_c, \phi \Sigma_c,$ and the $C_{ij}$ coefficients are given in table \ref{tab:vij1j13}  of the Appendix. The mixing of $PB$ and $VB$ states requites $\pi$ exchange, and
it was shown in Ref. \cite{liang} that this mixing was responsible for the breaking of the original degeneracy of two spin states of $\Lambda_b$, which upon  consideration of this mixing became the $\Lambda_b(5912)$ and $\Lambda_b(5920)$. We will discuss this mixing in Section \ref{sec:mixing}.

We also consider the singular case of single channel $\pi \Sigma_c^*$ in $I=0$ and $J=3/2$, where, as the $\pi \Sigma_b^*$ state in Ref. \cite{liang}, the mixing with $VB$ states that also gives $J=3/2$ with ordinary vector exchange in local hidden gauge approach is not allowed. We have the result for $C_{ij}$ coefficient given in Table \ref{tab:vij0j3} of the Appendix \cite{sarkar}.

\begin{figure}[tb]
\epsfig{file=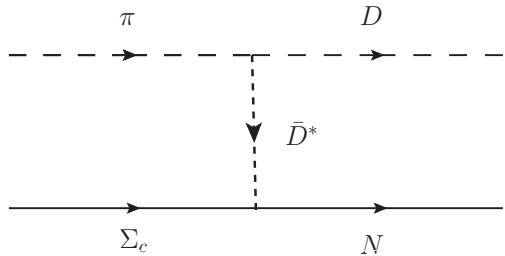, width=7cm}
\caption{Transition potential from $\pi \Sigma_c \to D N$.}
\label{fig:Bex}
\end{figure}

Before closing this section we must mention some feature concerning the transition $\pi \Sigma_c \to D N$. This is depicted in Fig. \ref{fig:Bex}.
and it is mediated by $\bar D^*$ exchange in the extended local hidden gauge approach. If we followed the strict large heavy quark mass counting we would neglect this term because it involves the exchange of a heavy vector $\bar D^*$ and its propagator would make this term small. However, although the term is suppressed, it is not as  much as one would expect. Indeed, the propagator will be
\begin{equation}
G_{D^*} = \frac{1}{p^2_{D^*} - m^2_{D^*}} \equiv \frac{1}{(p^0_{\pi} - p^0_{D})^2 - (\vec{p}_{\pi}-\vec{p}_{D})^2 - m^2_{D*}}.
\end{equation}
Conversely, in a diagonal transition $D N \to D N$ mediated by $\rho$ exchange, for instance, we have
\begin{equation}
G_\rho \approx \frac{1}{p_{\rho}^2 -m_V^2}.
\end{equation}
However, in this case and not far from the threshold, $p_{\rho}^2$ is small and is neglected (this approximation is implicit in the chiral Lagrangians).
Thus, close to $D N$ threshold the ratio is
\begin{equation}
\frac{G_{D^*}}{G_\rho} \simeq \frac{-m_V^2}{(p^0_{\pi} - p^0_{D})^2 - \vec{p}_{\pi}^2 - m^2_{D^*}} \simeq \frac{1}{4}.
\label{eq:approximation}
\end{equation}
This ratio was already noticed in Refs. \cite{angelsmizu, liang}. In Ref. \cite{Wu:2010vk}, the heavy meson propagators were explicitly used, and this approximation was also found fair.
Since the non diagonal terms have a smaller importance in the process than the diagonal ones of the heavy mesons, we simply account for these transitions multiplying by $1/4$ the results obtained from Eq. \eqref{eq:vij} and the Tables \ref{tab:vij0}-\ref{tab:vij0j3}.
Note that small variations from Eq. \eqref{eq:approximation} can be easily accommodated by means of changes in the cut off, that will be tuned to fit some data.

\subsection{Vector-baryon channels}

The transitions $VB \to VB$ for small three-momenta of the external vector mesons have formally the same expressions as the corresponding $PB \to PB$ substituting the octet of pseudoscalars by the octet of vectors \cite{angelsvec}. There is only one minor change needed to account for the $\phi$ and $\omega$ SU(3) structure, which is to replace each $\eta$ by $-\sqrt{2/3} ~\phi$ or $\sqrt{1/3} ~\omega$. The results for the $C_{ij}$ coefficients are given in Tables \ref{tab:vij0j13} and \ref{tab:vij1j13} of the Apendix.
Once again we suppress with a factor $1/4$ the transitions from a heavy vector to a light vector as done before for the pseudoscalar mesons.

The other novelty is that both the potential $V$ and the scattering matrix $T$ have the extra factor $ \vec{\epsilon} \cdot  {\vec{\epsilon}}~'$, where $\vec{\epsilon}$, ${\vec{\epsilon}}~'$ are the polarization vectors of the incoming and outgoing vector mesons. With this interaction there is degeneracy for spin states with $J=1/2$ and $J=3/2$ in $L=0$, which the only orbital angular momentum that we consider in the interaction of the channels.

\subsection{The construction of the scattering matrix}
In the former points, we have indicated how we construct the transition potential $V_{ij}$, or $V$ in matrix form. Next, we proceed to evaluate the scattering matrix $T$.

In coupled channels we make use of the Bethe-Salpeter equation
\begin{equation}
T = [1 - V \, G]^{-1}\, V,
\label{eq:Bethe}
\end{equation}
with $G$ the diagonal loop function for the propagating intermediate meson-baryon channels. In Ref. \cite{xiaooset} a warning was raised about potential dangers of using the dimensional regularization for the $G$ functions (see also Ref. \cite{wuzou}). This was so because for values of the energy below threshold, $G$ can become positive and then one can obtain bound states with a positive (repulsive) potential when $1 - VG =0$, which is not physically acceptable (see Eq. \eqref{eq:Bethe} in one channel). Hence, we use here the cut off regularization for $G$ given by
\begin{equation}
G(s) = \int_0^{q_{max}} \frac{d^{3}\vec{q}}{(2\pi)^{3}}\frac{\omega_P+\omega_B}{2\omega_P\omega_B}\,\frac{2M_{B}}{P^{0\,2}-(\omega_P+\omega_B)^2+i\varepsilon},
\label{eq:Gco}
\end{equation}
where $P^0$ is the CM energy, $s=(P^0)^2$, $\omega_P = \sqrt{\vec{q}\,^2+m_P^2},~\omega_B = \sqrt{\vec{q}\,^2+M_B^2}$, and $q_{max}$ is the cut-off of the three-momentum.\footnote{The cut off method should be applied in the Center of Mass (CM) frame of the two interacting particles. To guarantee Lorentz invariance, if we have the pair of particles in a moving frame, a boost must be done to the CM frame and evaluate there the scattering matrix.}

\section{Breaking the $J^P=1/2^-, \, 3/2^-$ degeneracy in the $D^* N$ sector}
\label{sec:mixing}

\begin{figure}[tb]
\epsfig{file=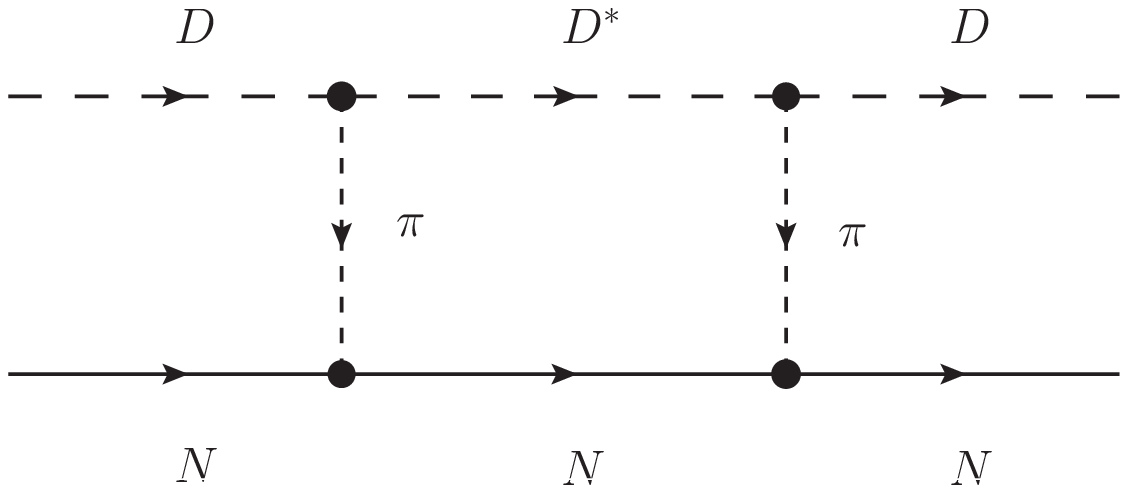, width=6.5cm} ~~~\epsfig{file=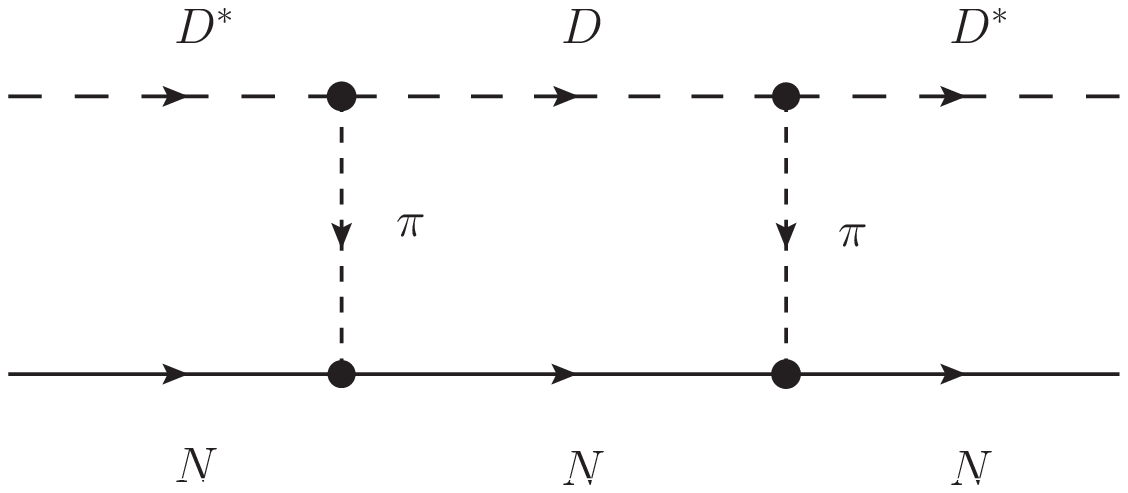, width=6.5cm}
\caption{Diagrammatic representation of the $D^* N$ in the intermediate state (left) and the $D N$ in the intermediate state (right).}\label{fig:bbbox}
\end{figure}

\begin{figure}[tb]
\epsfig{file=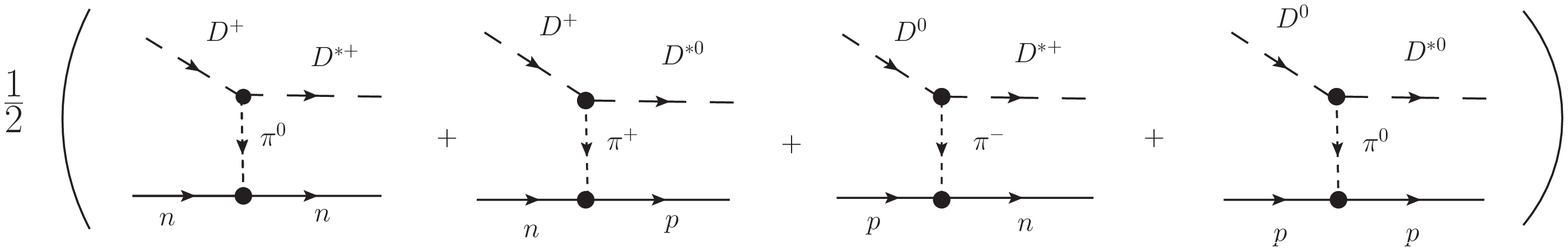, width=16cm}
\caption{Diagrammatic representation of the transition $D N \to D^* N$ in $I=0$.}\label{fig:bbboxhalf}
\end{figure}

In this section we break the degeneracy of the $J^P=1/2^-, \, 3/2^-$ states of the $D^* N$ sector. We follow the approach of Refs. \cite{javier,revhidden,kanchan,kanchan2} and mix states of $D^* N$ and $D N$ in both spin channels. We, thus, evaluate the contribution of the box diagrams of Fig. \ref{fig:bbbox},
in analogy to the box diagrams evaluated in Ref. \cite{javier}, and add this contribution, $\delta V$, to the $D N$ or $D^* N$ potential. Using the doublets of isospin $(\bar D^0, \, D^-)$, $(D^+, \, -D^0)$ the $\Lambda_c$ state in the $D N$ basis is given by
\begin{equation}
|D N, \, I=0 \rangle = \frac{1}{\sqrt{2}} (|D^+ n \rangle + |D^0 p \rangle),  \label{eq:dns}
\end{equation}
and analogously for $D^* N$. The $D N \to D^* N$ transition in $I=0$ is given by the diagrams of Fig. \ref{fig:bbboxhalf}.

The $V \, P \, \pi$ vertex in SU(3) is given by the Lagrangian
\begin{equation}
{\cal L}_{VPP} = -ig ~\langle [P,\partial_{\mu}P]V^{\mu}\rangle, \label{eq:vpp}
\end{equation}
where $P,\ V^\mu$ are the ordinary pseudoscalar octet and vector nonet SU(3) matrices of the corresponding fields
\begin{eqnarray}
P &=& \left(
\begin{array}{ccc}
\frac{\pi^0}{\sqrt{2}}+\frac{\eta_8}{\sqrt{6}}  & \pi^+     & K^{+}   \\
\pi^-      & -\frac{\pi^0}{\sqrt{2}}+\frac{\eta_8}{\sqrt{6}} & K^{0}   \\
K^{-}      & \bar{K}^{0}       & -\frac{2\eta_8}{\sqrt{6}}  \\
\end{array}
\right) \ ,  \\
V_\mu &=& \left(
\begin{array}{ccc}
\frac{\rho^0}{\sqrt{2}}+\frac{\omega}{\sqrt{2}} & \rho^+ & \quad K^{*+}  \\
 \rho^{-} & -\frac{\rho^0}{\sqrt{2}} + \frac{\omega}{\sqrt{2}} & K^{*0} \\
  K^{*-} & \bar K^{*0} & \phi \\
\end{array}
\right)_\mu \ .
\end{eqnarray}
and $g=m_V/2f_\pi$ with $m_V \approx 780 \mev$, $f_\pi=93$~MeV. One can extend the Lagrangian Eq. \eqref{eq:vpp} to the SU(4) space, as done in Ref. \cite{xiaoaltug}, but it is unnecessary. In Ref. \cite{liang} it was shown how to extend to the heavy sector the results of SU(3), using the impulse approximation at the quark level and considering the heavy quarks as spectators. The result obtained was that the $K^{*+} \to K^0 \pi^+$ and $B^{*+} \to B^0 \pi^+$ transition amplitudes were related by the relationship

\begin{equation}
\frac{t_{B^*}}{t_{K^*}} \equiv \frac{\sqrt{m_{B^*}m_B}}{\sqrt{m_{K^*}m_K}} \simeq \frac{m_{B^*}}{m_{K^*}}, \label{eq:ratio}
\end{equation}
the masses referring to the mesons.

In our case this ratio is changed to $m_{D^*}/m_{K^*}$. It was also found in Ref. \cite{liang} that the ratio of energies that must be implemented in the ratio of amplitudes for the  Weinberg-Tomozawa term was already provided by the form of the Weinberg-Tomozawa amplitude, Eq. (\ref{eq:vij}), that incorporates the energy of the mesons explicitly as a factor.

\begin{figure}[tb]
\epsfig{file=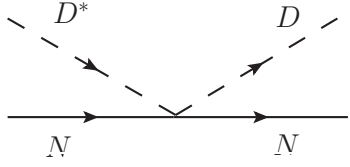, width=5.0cm}
\caption{Diagram of the Kroll-Ruderman term.}\label{fig:kroll}
\end{figure}

\begin{figure}[tb]
\epsfig{file=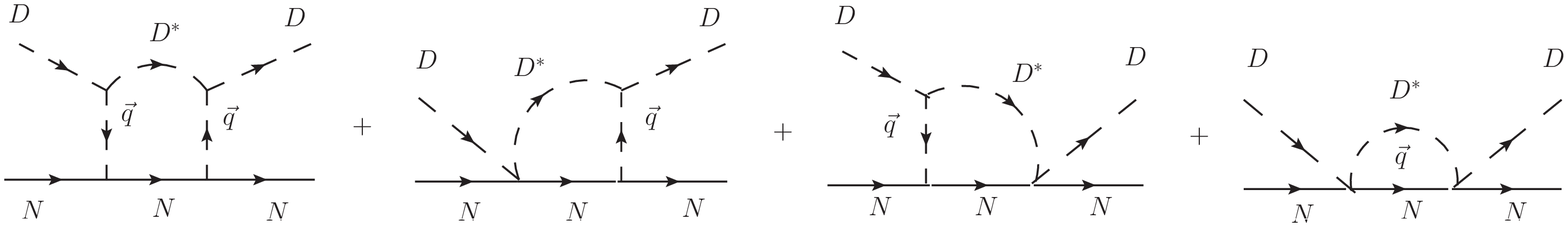, width=16cm}
\caption{All the diagrams for  $D^* N$ in the intermediate state.}\label{fig:bbboxtot}
\end{figure}

Now we come back to the evaluation of the box diagrams of Fig. \ref{fig:bbbox}. The vertex for the $I=0$ transition $D N \to D^* N$ of Fig. \ref{fig:bbboxhalf}, considering the Yukawa coupling for the $\pi NN$ vertex is given by
\begin{equation}
-it = - \frac{3}{\sqrt{2}} g \frac{m_{D^*}}{m_{K^*}} (q + P_{in})_\mu \epsilon^\mu \frac{1}{q^2 - m^2_\pi} \frac{D+F}{2f_\pi} \vec{\sigma}\,\cdot\, \vec{q}, \label{eq:tbox}
\end{equation}
with $D=0.75$ and $F=0.51$ \cite{Borasoy:1998pe}, and since $P_{in} = q + P_{out}$ and $P_{out} \cdot \epsilon =0$ plus $\epsilon^0 \approx 0$, we get effectively
\begin{equation}
-it = \frac{6}{\sqrt{2}} g \frac{m_{D^*}}{m_{K^*}} \vec{q}\,\cdot\, \vec{\epsilon} \frac{1}{q^2 - m^2_\pi} \frac{D+F}{2f_\pi} \vec{\sigma}\,\cdot\, \vec{q}.
\end{equation}
In addition to the pion exchange of Fig. \ref{fig:bbboxhalf}, we have the Kroll-Ruderman contact term, depicted in Fig. \ref{fig:kroll}.
Following Refs. \cite{javier,Carrasco:1989vq}, in order to get the Kroll-Ruderman term we must substitute in Eq. \eqref{eq:tbox} $\epsilon_\mu (q + P_{in})^\mu \frac{1}{q^2 - m^2_\pi} \vec{\sigma}\,\cdot\, \vec{q}$ by $-\vec{\sigma} \cdot \vec{\epsilon}$. Then, we must evaluate the diagrams of Fig. \ref{fig:bbboxtot}
and we obtain
\begin{equation}
\delta V = \delta V^{PP} + 2 \delta V^{PC} + \delta V^{CC}, \label{eq:delv}
\end{equation}
where $\delta V^{PP}$ stands for the first diagram of Fig. \ref{fig:bbboxtot}, $2 \delta V^{PC}$ for the two middle diagrams and $\delta V^{CC}$ for the last one.
The expressions for these terms can be seen in section V of Ref. \cite{liang}. All that is needed is to change the masses of $\bar B$ and $\bar B^*$ by those of $D$ and $D^*$. We write the expressions below.

We have for $DN \to D^* N \to DN$ box diagram
\begin{eqnarray}
-i\delta V^{PP} &=& \int \frac{d^4 q}{(2\pi)^4} \Big( \frac{m_{D^*}}{m_{K^*}} \Big)^2 \, g \big( \frac{6}{\sqrt{2}} \vec{\epsilon} \,\cdot\,\vec{q} \frac{1}{q^{0\,2} - \vec{q}\,^2 - m^2_\pi} \frac{D+F}{2f_\pi} \vec{\sigma}\,\cdot\, \vec{q} \big) \nonumber \\
&& \times (-g) \big( \frac{6}{\sqrt{2}} \vec{\epsilon} \,\cdot\,\vec{q} \frac{1}{q^{0\,2} - \vec{q}\,^2 - m^2_\pi} \frac{D+F}{2f_\pi} \vec{\sigma}\,\cdot\, \vec{q} \big) \nonumber \\
&& \times i \frac{1}{2 \omega_{D^*}(\vec{q}\,)} \frac{1}{P^0_{in}- q^0 - \omega_{D^*}(\vec{q}\,) + i \epsilon} i \frac{M_N}{E_N(\vec{q}\,)} \frac{1}{K^0_{in}+q^0 - E_N(\vec{q}\,) + i \epsilon}, \label{eq:delvpp} \\\nonumber
\end{eqnarray}
\begin{eqnarray}
-i\delta V^{PC} &=& \int \frac{d^4 q}{(2\pi)^4} \Big( \frac{m_{D^*}}{m_{K^*}} \Big)^2 \, g \big( \frac{3}{\sqrt{2}}  \frac{D+F}{2f_\pi} \vec{\sigma}\,\cdot\, \vec{\epsilon}  \big) (-g) \big( \frac{6}{\sqrt{2}} \vec{\epsilon} \,\cdot\,\vec{q} \frac{1}{q^{0\,2} - \vec{q}\,^2 - m^2_\pi} \frac{D+F}{2f_\pi} \vec{\sigma}\,\cdot\, \vec{q} \big) \nonumber \\
&& \times i \frac{1}{2 \omega_{D^*}(\vec{q}\,)} \frac{1}{P^0_{in}- q^0 - \omega_{D^*}(\vec{q}\,) + i \epsilon} i \frac{M_N}{E_N(\vec{q}\,)} \frac{1}{K^0_{in}+q^0 - E_N(\vec{q}\,) + i \epsilon}, \label{eq:delvpc} \\\nonumber
\end{eqnarray}
\begin{eqnarray}
-i\delta V^{CC} &=& \int \frac{d^4 q}{(2\pi)^4} \Big( \frac{m_{D^*}}{m_{K^*}} \Big)^2 \, g \big( \frac{3}{\sqrt{2}}  \frac{D+F}{2f_\pi} \vec{\sigma}\,\cdot\, \vec{\epsilon}  \big) (-g) \big( \frac{3}{\sqrt{2}}  \frac{D+F}{2f_\pi} \vec{\sigma}\,\cdot\, \vec{\epsilon}  \big) \nonumber \\
&& \times i \frac{1}{2 \omega_{D^*}(\vec{q}\,)} \frac{1}{P^0_{in}- q^0 - \omega_{D^*}(\vec{q}\,) + i \epsilon} i \frac{M_N}{E_N(\vec{q}\,)} \frac{1}{K^0_{in}+q^0 - E_N(\vec{q}\,) + i \epsilon},  \label{eq:delvcc}
\end{eqnarray}
where $P^0_{in}$ and $K^0_{in}$ are the energies of the incoming meson and baryon respectively, and $\omega_{D^*}(\vec{q})=\sqrt{\vec{q}^2 +m_{D^*}^2}$, etc., $E_N(\vec{q})=\sqrt{\vec{q}^2 +m_{N}^2}$.

\section{Box diagram for $I=1$ states}

We now evaluate the contribution of the box diagram to the $I=1$ states made from $D N$ and $D^* N$.

a) $D N$, $I=1$:

The isospin $I=1$ state is now
\begin{equation}
| D N; I=1, I_3=0 \rangle = \frac{1}{\sqrt{2}} \big( |D^+ n \rangle - | D^0 p \rangle \big).
\end{equation}
The counting of isospin done before for the $D^* \to D \pi$ transition can be repeated and we simply find that a factor $\frac{3}{\sqrt{2}}$ gets converted in $\frac{1}{\sqrt{2}}$ in the $D N \to D^* N$ transition. We thus get a factor 9 smaller contribution than for $I=0$ from the box and we neglect it.\\

b) $D^* N$, $I=1$:

We have the same suppression factor as before and we also neglect it. The same can be said with respect to the anomalous contribution.

As a consequence of the smallness of the $DN$ and $D^* N$ mixing in $I=1$, the two sectors appear differentiated and we do not perform the full coupled channel study in these cases.

\section{Contribution from the anomalous term }

\begin{figure}[tb]
\epsfig{file=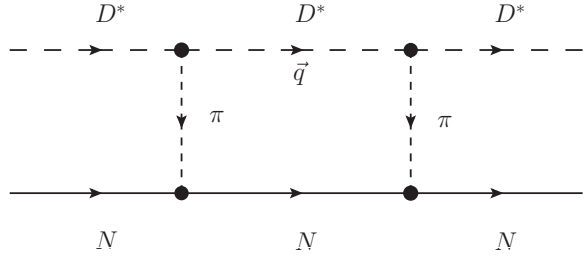, width=8cm}
\caption{Box diagram with anomalous $D^* D^* \pi$ vertex.}\label{fig:dsnano}
\end{figure}

For the $D^*N$ state it is possible to have an extra contribution, which does not interfere with the $s$-wave contribution at tree level. We take into account this contribution by means of the box diagram of Fig. \ref{fig:dsnano}.
The $I=0$ $D^*N$ state is given by Eq. \eqref{eq:dns} changing $D$ by $D^*$. The anomalous $D^* D^* \pi$ Lagrangian can be obtained from \footnote{An anomalous process, like the $VVP$ interaction, is one that does not conserve
``natural" parity.
The ``natural" parity of a particle is defined as follows: it is $+1$
if the particle transforms as a true tensor of that rank, and $-1$ if it
transforms as a pseudotensor, e.g. $\pi, \gamma, \rho$  and $a_1$ have ``natural"
parity $-1, +1, +1$ and $-1$, respectively.}
\begin{equation}
{\cal L}_{VVP} = \frac{G'}{\sqrt{2}} \epsilon^{\mu \nu \alpha \beta} \langle \partial_\mu V_\nu \partial_\alpha V_\beta P \rangle,
\end{equation}
with $G' = 3 M_V^2 /16 \pi^2 f^3_\pi (\simeq 14 \gev ^{-1})$ \cite{bramon,luisjose}. The $V$ matrix and $P$ matrix in SU(4) are given in Ref. \cite{danizou}. In addition we also need the standard Yukawa coupling of $\pi NN$, already considered in the evaluation of the former box diagrams. We evaluate this box diagram assuming that the three momentum of the $D^*$ is small compared with its mass and get for the anomalous $\pi$ exchange potential for $I=0$ $D^*N$ states
\begin{equation}
-it^{anom} = -\frac{1}{2} \epsilon^{ijk} \frac{G'}{\sqrt{2}} i \; m_{D^*} \; \epsilon_i (\vec{p})\; q_j \;\epsilon_k (\vec{q}\,) \; \vec{\sigma} \cdot \vec{q} \; \frac{F+D}{2f} \frac{G'}{\sqrt{2}} \frac{i}{q^2 - m^2_\pi}.
\end{equation}
The evaluation of the box diagram proceeds as in the former box diagram and we finally obtain
\begin{equation}
\delta V^{box}_{ano} = 3 \frac{G'^2}{2} \big( \frac{F+D}{2f} \big)^2 m_{D^*}^2 \frac{\partial I_1}{\partial m^2_\pi}  \frac{1}{4},
\end{equation}
with $I_1$ given by Eq. (41) of Ref. \cite{liang} changing the masses of $B, \ B^*$ by those of $D, \ D^*$.

We find that the contribution of $\delta V^{box}_{ano}$ is of the order of $\delta V$ in the box diagram of $D^* N \to DN \to D^* N$ and about half the size of $\delta V$ in the box diagram of $DN \to D^* N \to DN$.

\section{Full coupled channel calculation for $I=0$}

In Ref. \cite{liang}, the box diagrams contributions to the potential, $\delta V$ of  Eq. \eqref{eq:delv}, were added to the potential $V$ and the Bethe Salpeter equation was solved. While the procedure is fair when one has some dominant channel, like $BN, B^* N$ in Ref. \cite{liang}, in the case where there is a stronger mixture of channels, a full coupled channels calculation, where the $VB$ and $PB$ channels are treated on the same footing, becomes advisable. The study done in Ref. \cite{liang} shows that the $\Lambda_b(5912)$ and $\Lambda_b(5920)$ are basically $\bar B^* N$ states degenerate when only the $VB$ states are considered, and the degeneracy is broken when the box diagram with intermediate $\bar B N$ states is considered. In the present case, it was found in  Ref. \cite{GarciaRecio:2008dp} that the $\Lambda_c(2595)$ state had a large admixture of $DN$ and $D^* N$, with the $D^* N$ channel being dominant. In that case an SU(8) symmetry scheme was used to connect the $PB$ and $VB$ channels with no explicit reference to the dynamics of pion exchange linking these states. In the present  work, we undertake the task of performing the full coupled channels study of the $PB$ and $VB$ states using the explicit dynamics of pion exchange.
For this we have to look in detail into the expressions of $\delta V^{PP}$, $\delta V^{PC}$, $\delta V^{CC}$ of Eqs. \eqref{eq:delv}-\eqref{eq:delvcc}.

The first thing that we do is to separate the $s$-wave part from the $d$-wave part of the box diagram. This must be done because we only take into account the pseudoscalar-baryon interactions in $s$-wave. Then we keep the $d$-wave contribution of the box diagram as an additional contribution $\delta V (d$-wave) to the $DN \to DN$ or $D^* N \to D^* N$ potentials. To account for the $s$-wave part we consider the diagrams of Fig. \ref{fig:bbboxtot} and define an effective transition potential from $DN \to D^* N$ as
\begin{equation}
\tilde{V}^2_{eff} G_{D^* N} = \delta V_1 (s\textrm{-wave}). \label{eq:eff1}
\end{equation}
Certainly we can also define $\tilde{V}_{eff}$ from the diagrams equivalent to Fig. \ref{fig:bbboxtot} but for $D^* N$ having $DN$ as intermediate states. We would have then
\begin{equation}
\tilde{V}^{'\,2}_{eff} G_{D N} = \delta V_2 (s\textrm{-wave}). \label{eq:eff2}
\end{equation}
The idea is to use this transition potential $\tilde{V}_{eff}$ in coupled channels of $PB$ and $VB$ together. The definitions of Eqs. \eqref{eq:eff1} and \eqref{eq:eff2} guarantee that the effect of the box is accounted for, but in addition, the full coupled channels approach will take into account the interaction of $D^* N$ in intermediate states in Fig. \ref{fig:bbboxtot} or of $DN$ in the equivalent diagrams for the box diagrams of $D^* N$ with $DN$ intermediate states. Since in the coupled channels approach $V_{ij} = V_{ji}$, there is here the ambiguity that $\tilde{V}_{eff}$ and $\tilde{V}'_{eff}$ are not necessarily equal. Fortunately they turn out to be very similar (they are equal if $m_{D^*} = m_D$ as we shall see below), but in order to guarantee $V_{ij} = V_{ji}$ we take a unique $V_{eff}$ defined as
\begin{equation}
V_{eff} = \frac{1}{2} (\tilde{V}_{eff} + \tilde{V}'_{eff}). \label{eq:eff}
\end{equation}

The next step consists of evaluating explicitly the $s$-wave and $d$-wave parts of the box diagram:
\newline 1) The terms containing a Kroll-Ruderman vertex in Fig. \ref{fig:bbboxtot} (last three diagrams) contain a $\vec{\sigma} \cdot \vec{\epsilon}$ vertex which projects over $s$-wave automatically.
\newline 2) The double pion pole term (first diagram of Fig. \ref{fig:bbboxtot}) is split as
\begin{equation}
\vec{\epsilon} \cdot \vec{q} \quad \vec{\sigma} \cdot \vec{q} \to \epsilon_i q_i \sigma_j q_j = \epsilon_i \sigma_j \{ \frac{1}{3} \vec{q}\,^2 \delta_{ij} + (q_i q_j - \frac{1}{3}  \vec{q}\,^2 \delta_{ij}) \},
\end{equation}
where the first term in the last expression of the equation stands for the $s$-wave contribution and the second one for the $d$-wave. The $s$-wave contribution gives
\begin{equation}
\frac{1}{3} \vec{q}\,^2 \vec{\sigma} \cdot \vec{\epsilon} = \frac{1}{3} \vec{q}\,^2 \sqrt{3},
\end{equation}
where we have used that $\langle PN| \vec{\sigma} \cdot \vec{\epsilon} |VN \rangle = \sqrt{3} \delta_{J,1/2}$, from Ref. \cite{javier}, and for the two $\pi$ exchanges in the box diagram we get
\begin{equation}
s\textrm{-wave}: (\frac{1}{3} \vec{q}\,^2 \vec{\sigma} \cdot \vec{\epsilon}\,)^2 = \frac{1}{3} \vec{q}\,^4.
\end{equation}
The $d$-wave is better evaluated subtracting this amount from the total. The total contribution is obtained using
\begin{eqnarray}
&& \vec{\sigma} \cdot \vec{q} \quad \vec{\sigma} \cdot \vec{q} \to \vec{q}\,^2; \label{eq:sigma2} \\
&& \vec{\epsilon} \cdot \vec{q} \quad \vec{\epsilon} \cdot \vec{q} \to \epsilon_i \epsilon_j q_i q_j \to \sum_{\lambda} \epsilon_i^{(\lambda)} \epsilon_j^{(\lambda)} \frac{1}{3} \vec{q}\,^2 \delta_{ij} = \frac{1}{3} \vec{q}\,^2 \delta_{ij} \delta_{ij} = \vec{q}\,^2,\label{eq:epsilon2}
\end{eqnarray}
where we have summed over the polarization of the intermediate vector states for $DN \to D^* N \to DN$. Altogether from Eqs. \eqref{eq:sigma2} and \eqref{eq:epsilon2}, we find $\vec{q}\,^4$ for this term and thus the $d$-wave contribution is
\begin{equation}
d\textrm{-wave}:  \frac{2}{3} \vec{q}\,^4.
\end{equation}

The former separation allows us to use the results of Ref. \cite{liang} and immediately write in analogy to Eqs. (40)-(43) of Ref. \cite{liang},
\begin{eqnarray}
&& \delta V (d\textrm{-wave}, DN \to D^* N \to DN) \approx \frac{2}{3} FAC \frac{\partial}{\partial m^2_\pi} I_1, \\
 \delta V_1 (s\textrm{-wave})&=& \delta V (s\textrm{-wave}, DN \to D^* N \to DN) \approx FAC ( \frac{1}{3} \frac{\partial}{\partial m^2_\pi} I_1 + 2 I_2 + I_3),
\label{eq:deltaV1}
\end{eqnarray}
which holds for $J=1/2$ in this case since $DN$ is in $s$-wave.
\newline 3) For the case of $D^* N \to DN \to D^* N$, if we have $J=3/2$, this forces the $DN$ state to be in $d$-wave, and the contribution of the box diagram does not have $s$-wave component in $DN$.  Then we have (see Eq. (33) of Ref. \cite{liang})
\begin{eqnarray}
\delta V (D^*N \to DN \to D^*N, J=3/2) = FAC \frac{\partial}{\partial m^2_\pi} I^{'}_1,
\end{eqnarray}
and we add it to the $D^*N \to D^*N$ potential.

If $J=1/2$ then the $DN$ intermediate state is necessarily in $s$-wave and we have (see Eq. (32) of Ref. \cite{liang})
\begin{eqnarray}
 \delta V_2 (s\textrm{-wave})= \delta V (D^*N \to DN \to D^*N, J=1/2) = FAC ( \frac{\partial}{\partial m^2_\pi} I^{'}_1 + 2I^{'}_2 + I^{'}_3 ).
 \label{eq:deltaV2}
\end{eqnarray}
With Eqs. (\ref{eq:eff1})-(\ref{eq:eff}), (\ref{eq:deltaV1}) and (\ref{eq:deltaV2}), one can get the effective transition potential $V_{eff}$. The expressions of $I^{'}_1, I^{'}_2, I^{'}_3, FAC, I_1, I_2,$ and $I_3$ can be seen in Eqs. (34)-(37) and (41)-(43) of Ref. \cite{liang},
changing the masses of $\bar B$ and $\bar B^*$ by those of $D$ and $D^*$.

It is interesting to note that formally $\tilde{V}_{eff}$ and $\tilde{V}'_{eff}$ are the same since $\frac{1}{3} I_1$ in Eq. (41) of Ref. \cite{liang} and $I'_1$ of Eq. (34) of Ref. \cite{liang} are equivalent and only the different masses of $D$, $D^*$ in the intermediate states make the numerical results different, however, not so much as pointed above, such that the concept of $V_{eff}$ for the transition is well defined. The space of states for $J=1/2$ is now $DN$, $\pi \Sigma_c$, $\eta \Lambda_c$, $D^* N$, $\rho \Sigma_c$, $\omega \Lambda_c$, $\phi \Lambda_c$, and for $J=3/2$, $D^* N$, $\rho \Sigma_c$, $\omega \Lambda_c$, $\phi \Lambda_c$.

\section{Results with full coupled channels for $I=0$}

In this section we show the results of using $DN$ and $D^*N$ as coupled channels in the way described in the formers sections. We shall take into account the freedom that we have in the different loops, $PB$ loop, $VB$ loop, box diagram, to take three different cutoffs and adjust them to a few experimental data. Then we proceed as follows. First, we take the $J^P = 3/2^-$ sector. The states are now $D^* N$, $\rho \Sigma_c$, $\omega \Lambda_c$, $\phi \Lambda_c$, since $DN$ does not couple to $J=3/2$ in $L=0$. In this case we have only two independent cutoffs, $q_{max}^V$ for the $D^* N$ loop and $q_{max}^B$ for the cutoff in the box diagrams including the box with the anomalous vertex. Since $q_{max}^B$ will be adjusted, we take $\Lambda$ of the Yukawa form factor, $F(\vec{q}\,) = \Lambda^2 / (\Lambda^2 + \vec{q}\,^2)$, in the $\pi B B'$ vertices as $\Lambda = 1000 \mev$, a standard value. Then we take values for $q_{max}^V$ similar to those found in the literature around 800 MeV \cite{wuzou} and change $q_{max}^B$ accordingly such as to get the energy of $2628 \mev$ for the bound $D^* N$ state. We get a range of values for both cutoffs as shown in Table \ref{tab:poleDsNAnomI0}.
\begin{table}[H]
     \renewcommand{\arraystretch}{1.5}
     \setlength{\tabcolsep}{0.2cm}
     \caption{Poles in coupled channel $D^*N[2946], \rho \Sigma_c[3229], \omega \Lambda_c [3066], \phi \Lambda_c[3306]$ with $I=0$ and $J=3/2$ with the anomalous term as a function of $q_{max}^{B,V}$. (The number in brackets after the channel indicates the mass of the channel. Units: MeV)}
\label{tab:poleDsNAnomI0}
\centering
   \begin{tabular}{|c|c|c|c|}
   \hline
   $q_{max}^B$ & 600 & 800 & 1000 \\
   $q_{max}^V$ & 771 & 737 &  715 \\
   \hline
                & $2628.45$ & $2628.35$ & $2628.27$   \\
   \cline{2-4}  & $2969.64+i0.91$ & $2990.43+i0.81$ & $3004.21+i0.77$ \\
   \hline
   \end{tabular}
\end{table}
We see that for all the combinations of $q_{max}^B$, $q_{max}^V$ we get the same value of the energy of the lower state (with zero width), but we also get a second state at energies around $2990 \mev$ with a very small width of $\Gamma < 2 \mev$. We observe, however, that the value of the energy varies from $2969 \mev$ to $3004 \mev$. So, we can accept these differences as uncertainties and state that we get an energy of about $2990 \mev$ with $\pm 20 \mev$ uncertainty. We think that the range given to $q_{max}^B$ is sufficiently wide to account for a fair uncertainty in the obtained masses.

It is most instructive to see the nature of the states, which we find by looking both at the couplings of the resonance to the different coupled channels, or by looking at the wave functions at the origin $g_i G_i^{II}$, with the coupling and $G$ functions evaluated at the pole in the second Riemann sheet. We can see the results in Table  \ref{tab:giDsNAnomI0b} for the middle set of cut off parameters in Table  \ref{tab:poleDsNAnomI0}. We have seen that the results are remarkably similar using the other sets of cut offs.
We can see in this table that both the couplings and the wave functions at the origin indicate that the state at $2628 \mev$ is essentially a $D^* N$ state, with a small admixture of the other channels.

\begin{table}[htp]
     \renewcommand{\arraystretch}{1.2}
     \caption{The coupling constants to various channels for the poles in the $I=0, \ J^P=3/2^-$ sector of $D^* N$ and coupled channels, with the anomalous term and taking $q_{max}^{B,V}=800,737$ MeV.  In bold face we highlight the main components.}
\label{tab:giDsNAnomI0b}
\centering
\begin{tabular}{ccccc}
\hline\hline
   $2628.35$ & $ D^*N $ & $\rho \Sigma_c$ & $\omega \Lambda_c$ &$\phi \Lambda_c$ \\
   \hline
   $g_i$  & $\bf 10.11$ & $-0.55$ & $0.49$ & $-0.68$ \\
   $g_i\,G_i^{II}$ & $\bf -29.10$ & $2.60$ &$-2.78$& $2.50$ \\
   \hline
   $2990.43+i0.81$ & $ D^*N $ & $\rho \Sigma_c$ & $\omega \Lambda_c$ &$\phi \Lambda_c$ \\
   \hline
   $g_i$                & $0.06+i0.11$ & $\bf 5.44+i0.01$ & $0.03+i0.02$ & $-0.04-i0.03$ \\
   $g_i\,G_i^{II}$ & $-1.23-i0.79$ & $\bf -44.53-i0.15$ & $-0.39-i0.25$ & $0.25+i0.16$ \\
   \hline
\end{tabular}
\end{table}

The state at $2990 \mev$ couples most strongly to $\rho \Sigma_c$ and, once again, the values of the couplings and the wave functions at the origin do not change much for one set of values of the cut offs to the other.

Next we turn to the states with $J^P = 1/2^-$. We can get them in $L=0$ from the $DN, \ D^* N$ and the different $PB$ and $VB$ states. Thus, the coupled channels are now $DN$, $\pi \Sigma_c$, $\eta \Lambda_c$, $D^* N$, $\rho \Sigma_c$, $\omega \Lambda_c$, $\phi \Lambda_c$. The strategy that we follow now is that we take $q_{max}^B$, $q_{max}^V$ as was obtained before for $J=3/2$. Then we fit the only free parameter $q_{max}^P$ in order to get the mass of the $J=1/2$, $\Lambda_c^* (2595)$ state at $2592 \mev$. We use the three sets of $q_{max}^B$, $q_{max}^V$ and show the results in Table \ref{tab:poleDNAnomI0}.
The values of $q_{max}^P$ are around $500 \mev$, in tune with the value $630 \mev$ that was used in the study of the $\bar{K} N$ and coupled channels system \cite{angels}. We find in the table four states.
\begin{table}[H]
     \renewcommand{\arraystretch}{1.5}
     \setlength{\tabcolsep}{0.2cm}
     \caption{Poles in coupled channel $DN[2808]$, $\pi \Sigma_c[2592]$, $\eta \Lambda_c[2834]$, $D^*N[2946]$, $\rho \Sigma_c[3229]$, $\omega \Lambda_c [3066]$, $\phi \Lambda_c[3306]$ with $I=0$ and $J=1/2$, with the anomalous term, as a function of $q_{max}^{B,V,P}$. (The number in brackets after the channel indicates the mass of the channel. Units: MeV)}
     \label{tab:poleDNAnomI0}
\centering
   \begin{tabular}{|c|c|c|c|}
	   \hline
   $q_{max}^B$ & 600 & 800 &  1000 \\
   $q_{max}^V$ & 771 & 737 &  715  \\
   $q_{max}^P$ & 527 & 500 &  483 \\
   \hline
                & $2592.26+i0.56$  & $2592.24+i0.55$ & $2592.14+i0.52$   \\
   \cline{2-4}  & $2610.44+i48.68$ & $2611.06+i53.35$ & $2611.32+i56.28$ \\
   \cline{2-4}  & $2757.25+i1.20$ & $2767.14+i0.98$ & $2772.41+i0.86$ \\
   \cline{2-4}  & $2970.01+i0.45$ & $2990.78+i0.60$ & $3004.54+i0.64$ \\
   \hline
   \end{tabular}
\end{table}

In Table \ref{tab:giJ3/2AnomI0b}, we show the couplings and wave functions at the origin of the different states to the coupled channels for the middle set of parameters of Table \ref{tab:poleDNAnomI0}. The results do not change much from one set of cut offs to another. We observe that the state at $2592 \mev$, which comes with a width of around $1 \mev$, and hence is compatible with experiment, couples strongly both to $DN$ and $D^* N$, and the couplings have opposite sign. We observe that the coupling to $\pi \Sigma_c$ is small, however, the wave function at the origin is relatively large. This can be explained because the energy 2592 MeV is just at threshold of $\pi \Sigma_c$. Then the modulus of the $G_i$ function has its maximum precisely at the threshold of channel $i$, and the value of $g_i G_i$ gets enhanced. The new state that has emerged at $2767 \mev$ also couples strongly to $DN$ and $D^* N$, but the couplings now have the same sign. Roughly speaking we have obtained two orthogonal states of $DN$, $D^* N$ of the type $\frac{1}{\sqrt{2}} (DN \pm D^* N)$. We also see that the state at $2611 \mev$ couples mostly to $\pi \Sigma_c$ and the one at $2990 \mev$ to $\rho \Sigma_c$. The $2990 \mev$ state with $J=3/2$ obtained in Table \ref{tab:poleDsNAnomI0} is practically degenerate with the one with $J=1/2$ which also couples mostly to $\rho \Sigma_c$.

An interesting thing of our approach is that even fixing the parameters to get the $2592 \mev$ and $2628 \mev$ states, we still have some freedom in the parameters and we see that there can be a fluctuation of about $\pm 10 \mev$ with respect to the central value for the $2767 \mev$ state and $\pm 20 \mev$ for the state with $2990 \mev$.

\begin{table}[htp]
     \renewcommand{\arraystretch}{1.2}
     \caption{The coupling constants to various channels for the poles in the $I=0, J^P=1/2^-$ sector, with the anomalous term and taking $q_{max}^{B,V,P}=800,737,500$ MeV. In bold face we highlight the main components.}
     \label{tab:giJ3/2AnomI0b}
\centering
\begin{tabular}{ccccc}
\hline\hline
   $2592.26+i0.56$& $DN $ & $\pi \Sigma_c$ & $\eta \Lambda_c$ &  \\ 
   \hline
   $g_i$  & $\bf -8.18+i0.61$ & $\bf 0.54+i0.00$ & $-0.40-i0.03$ & \\
   $g_i\,G_i^{II}$ & $\bf 13.88-i1.06$ & $\bf -10.30-i0.69$ & $1.76-i0.14$ & \\
   \hline
    & $ D^*N $ & $\rho \Sigma_c$ & $\omega \Lambda_c$ &$\phi \Lambda_c$ \\
   \hline
   $g_i$  & $\bf 9.81+i0.77$ & $-0.45-i0.04$ & $0.42+i0.03$ & $-0.59-i0.05$ \\
   $g_i\,G_i^{II}$ & $\bf -26.51-i2.10$ & $2.07+i0.17$ & $-2.31-i0.19$ & $2.10+i0.17$ \\
   \hline
   $2611.06+i53.35$ & $ DN $ & $\pi \Sigma_c$ & $\eta \Lambda_c$ &  \\ 
   \hline
   $g_i$  & $0.08-i1.81$ & $\bf 1.78+i1.40$ & $0.03-i0.09$ & \\
   $g_i\,G_i^{II}$ & $-0.68+i3.13$ & $\bf -55.22-i18.22$ & $-0.18+i0.39$ & \\
   \hline
    & $ D^*N $ & $\rho \Sigma_c$ & $\omega \Lambda_c$ &$\phi \Lambda_c$ \\
   \hline
   $g_i$  & $ -1.56+i1.38$ & $0.09-i0.05$ & $-0.08+i0.05$ & $0.11-i0.07$ \\
   $g_i\,G_i^{II}$ & $ 4.66-i3.42$ & $-0.44+i0.20$ & $0.46-i0.25$ & $-0.42+i0.24$ \\
   \hline
   $2767.14+i0.98$& $ DN $ & $\pi \Sigma_c$ & $\eta \Lambda_c$ &  \\ 
   \hline
   $g_i$  & $\bf -3.70+i0.04$ & $0.02-i0.20$ & $-0.52+i0.00$ & \\
   $g_i\,G_i^{II}$ & $\bf 14.78-i0.05$ & $3.54+i2.76$ & $4.40+i0.02$ & \\
   \hline
    & $ D^*N $ & $\rho \Sigma_c$ & $\omega \Lambda_c$ &$\phi \Lambda_c$ \\
   \hline
   $g_i$  & $\bf -3.97+i0.05$ & $0.47-i0.00$ & $-0.30+i0.00$ & $0.43-i0.00$ \\
   $g_i\,G_i^{II}$ & $\bf 15.47-i0.16$ & $-2.62+i0.01$ & $2.16-i0.02$ & $-1.82+i0.02$ \\
   \hline
   $2990.78+i0.60$& $ DN $ & $\pi \Sigma_c$ & $\eta \Lambda_c$ &  \\ 
   \hline
   $g_i$  & $0.01+i0.00$ & $0.00+i0.00$ & $-0.00-i0.01$ & \\
   $g_i\,G_i^{II}$ & $0.09+i0.14$ & $0.01+i0.03$ & $0.16-i0.08$ & \\
   \hline
    & $ D^*N $ & $\rho \Sigma_c$ & $\omega \Lambda_c$ &$\phi \Lambda_c$ \\
   \hline
   $g_i$  & $-0.09-i0.11$ & $\bf -5.44-i0.02$ & $-0.04-i0.01$ & $0.05+i0.02$ \\
   $g_i\,G_i^{II}$ & $1.57+i0.59$ & $\bf 44.54+i0.20$ & $0.50+i0.19$ & $-0.32-i0.12$ \\
   \hline
\end{tabular}
\end{table}

  It is instructive to compare the results obtained here with those of Ref. \cite{Romanets:2012hm}. There a $J^P$ $=1/2^-$, $I=0$ state is found at 2618 MeV, with very small width, which is associated to the $\Lambda_c(2595)$. The state couples both to $DN$ and $D^* N$ but the coupling to $D^*N$ is about 60 \% bigger than the one to $DN$. In our case the ratio of couplings to $D^* N$ and $DN$ is of the order of 1.20. It is also interesting to note that in  Ref. \cite{Romanets:2012hm} another $\Lambda_c$ resonance is found around 2617 MeV, but with a width of 90 MeV.  We also find a similar state around 2611 MeV and a width of 106 MeV.  In both cases, a considerable coupling to the $\pi \Sigma_c$ state is responsible for the width. In addition, in Ref. \cite{Romanets:2012hm} a state with $J^P=3/2^-$, $I=0$ is obtained at 2666 MeV with a width of 54 MeV which is associated to the $\Lambda_c(2625)$. We, instead, get a state at 2628 MeV and with zero width, with the dominant coupling to the $D^* N$ state, while in  Ref. \cite{Romanets:2012hm} there is a large coupling to $D^* N$ but there is also some coupling to $\pi \Sigma_c^*$ which is responsible for the relatively large width. Yet, the width can be drastically reduced if the mass goes down, getting closer to the threshold mass of the $\pi \Sigma_c^*$ channel (2656 MeV). The dynamics of our approach highly suppresses this latter channel, which would involve $D$ exchange instead of $\pi$ exchange, and is hence further suppressed than the already suppressed pion exchange. Also in the  $J^P=1/2^-$, $I=0$ sector, in Ref. \cite{Romanets:2012hm} a state with 2828 MeV and a width of 0.8 MeV is found, which couples mostly to $\rho \Sigma_c$ among other channels.  We find a similar state at 2990 MeV of dominant $\rho \Sigma_c$ nature, with a width of about 2 MeV.

The result obtained here for the $\Lambda_c(2595)$ also agrees qualitatively with the one in Ref. \cite{angelsmizu}, or Ref. \cite{Lutz:2005vx}, where also the Weinberg-Tomozawa interaction is used, with some small differences in the coupling constants and the use of extra channels in Ref. \cite{angelsmizu} which are farther away in energy and which we have ignored. The main difference is that in these works, the state associated to the $\Lambda_c(2595)$ couples mostly to $D N$, while in our case it appears as a mixture of $D^* N$ and $DN$, more closely to the results of Ref. \cite{Romanets:2012hm}. It is interesting to see what happens in Ref. \cite{Romanets:2012hm} to the orthogonal state to the one at $2592 \mev$ in the $D^* N$ and $DN$ mixing. Our orthogonal combination leads in our case to the $2767 \mev$ state, while in the approach of Ref. \cite{Romanets:2012hm} this orthogonal combination has a zero eigenvalue for the binding \cite{troia}. In our case we get a state bound by about $40 \mev$ with respect to the $DN$ threshold, which is small compared with the about $200 \mev$ binding in the case of the $2592 \mev$ state. The qualitative agreement in the two schemes, which look quite different, is remarkable.

 In addition, the results for the single channel $\pi \Sigma^*_c$ in $I=0$ and $J=3/2$ are shown in Table \ref{tab:polePBSingleI0}.

 \begin{table}[htp]
     \renewcommand{\arraystretch}{1.5}
     \setlength{\tabcolsep}{0.2cm}
     \caption{Poles in single channel $\pi \Sigma^*_c$[2656] with $I=0$ and $J=3/2$ as a function of $q_{max}^P$. (The number in brackets after the channel indicates the mass of the channel. Units: MeV)}
\label{tab:polePBSingleI0}
\centering
\begin{tabular}{|c|c|c|c|}
\hline
$q_{max}^P$ & 527 & 500 & 483 \\
\hline
 no box &  $2673.14 +i51.55$  &  $2673.95 + i55.63$  &  $2674.34 + i58.25$ \\
\hline
\end{tabular}
\end{table}

 By looking at table III of Ref. \cite{Romanets:2012hm} this state is likely to be identified with the $J=3/2$ state at 2666 MeV of Ref. \cite{Romanets:2012hm} which couples strongly to $\pi \Sigma_c^*$ and was associated there to the experimental state at 2628 MeV.  In our case the state associated to the experimental $J=3/2$ state $\Lambda_c(2625)$ has a different nature and is mostly a $D^*N$ state. If we force the state in Table \ref{tab:polePBSingleI0} to correspond to the experimental one with $J=3/2$, we need $q_{max}=1530$ MeV, which we could not justify with the ranges found from phenomenology.
 We stick to the choice of the cut off $q_{max}^P=500$ MeV for the $PB$ channels in which case we have a prediction of a state with $I=0,~J=3/2$ of 2674 MeV with $\Gamma=111$ MeV. We should note that we have obtained a resonant state above threshold with a single channel. This might seem to contradict the findings in Ref. \cite{yama} where a single channel with an energy independent potential does not generate resonances above threshold. We have checked that it is the energy dependence of the potential of Eq. (\ref{eq:vij}) what makes the appearance of the state possible. Indeed, if we make the potential energy independent by taking its value at the $\pi \Sigma_c^*$ threshold we do not get poles above threshold.

\section{$I=1$ states}

In this section we show the results that we obtain for $I=1$ from the $DN$, $D^* N$, and coupled channels. We already discussed that the box diagram for the case of the $I=1$ $DN$ and $D^* N$ states is negligible. In this case we do not need to use the full coupled channel approach and we have separated coupled channels for $PB$ and $VB$. We use the same cut offs obtained before.

In Table \ref{tab:poleDNCoupI1} we show the pole position of the state obtained with the $DN$ and coupled channels as a function of the cut off.

\begin{table}[H]
     \renewcommand{\arraystretch}{1.5}
     \setlength{\tabcolsep}{0.2cm}
     \caption{Poles in the $I=1$ sector of $D N$ and coupled channels as a function of $q_{max}^P$. Threshold masses in brackets: $DN$[2806], $\pi \Sigma_c$[2592], $\pi \Lambda_c$[2425],  $\eta \Sigma_c$[3001]. (Units: MeV)}
\label{tab:poleDNCoupI1}
\centering
\begin{tabular}{|c|c|c|c|}
\hline
$q_{max}^P$ & 527  & 500  & 483  \\
\hline
no box  &  $2668.92 +i131.63$  & $2665.81 + i136.10$  & $2664.89 +i138.83$   \\
\hline
\end{tabular}
\end{table}

  By taking the pole obtained for $q_{max}^P= 500$  MeV we show in Table \ref{tab:giDNI1} the couplings and wave functions at the origin. We observe that the state largely couples to $\pi \Sigma_c$. The couplings and wave functions at the origin with the other sets of parameters are basically the same.

\begin{table}[ht]
     \renewcommand{\arraystretch}{1.2}
     \caption{The coupling constants to various channels for the poles in the $I=1$ sector of $D N$ and coupled channels, taking $q_{max}^P=500$ MeV.  In bold face we highlight the main component.} \label{tab:giDNI1}
\centering
\begin{tabular}{ccccc}
\hline\hline
$2665.81 +i136.10$ & $DN$ & $\pi \Sigma_c$ & $\pi \Lambda_c$ & $\eta \Sigma_c$ \\
\hline
$g_i$ & $-1.07-i0.58$~~ & $\bf 1.55 +i1.55$ & $0.02+ i0.04$  & $0.02+ i0.04$  \\
$g_i\,G_i^{II}$ & $1.26 + i1.92$~~ & $\bf -72.39- i15.30$~~ & $-1.57- i0.10$~~ & $-0.02 -i0.13$~~  \\
\hline
\end{tabular}
\end{table}

In Table \ref{tab:poleDSNCoupI1} we show the states obtained with  $D^*N$ and its coupled channels as a function of the cut off. We find two states with zero or a small width. The couplings of these states to the coupled channels are shown in Table \ref{tab:giDsNI1} for the cut off that we used with the same $D^*N$ channel in $I=0$. We see that the state found around 2929 MeV couples mostly to $D^* N$, while the one at 3146 MeV couples mostly to $\rho \Sigma_c$.

\begin{table}[H]
\renewcommand{\arraystretch}{1.5}
\setlength{\tabcolsep}{0.2cm}
\caption{Poles in the $I=1$ sector of $D^* N$ and coupled channels as a function of $q_{max}^V$. Threshold masses in brackets: $D^* N$[2948], $\rho \Sigma_c$[3229], $\rho \Lambda_c$[3062], $\omega \Sigma_c$[3236], $\phi \Sigma_c$[3473]. (Units: MeV)}
\label{tab:poleDSNCoupI1}
\centering
\begin{tabular}{|c|c|c|c|}
\hline
$q_{max}^V$ & 771  & 737  & 715\\
\hline
  & $2922.09+i0$  & $2928.85+i0$  & $2932.66+i0$\\
\cline{2-4} \rb{no box} & $3133.27+i4.07$  & $3145.71 + i3.57$ & $3153.51 +i3.22$ \\
\hline
\end{tabular}
\end{table}

\begin{table}[ht]
     \renewcommand{\arraystretch}{1.2}
     \caption{The coupling constants to various channels for the poles in the $I=1$ sector of $D^* N$ and coupled channels, taking $q_{max}^V=737$ MeV. In bold face we highlight the main components.} \label{tab:giDsNI1}
\centering
\begin{tabular}{cccccc}
\hline\hline
$2928.85+i0$ & $D^* N$ & $\rho \Sigma_c$ & $\rho \Lambda_c$ & $\omega \Sigma_c$ & $\phi \Sigma_c$\\
\hline
$g_i$ & ~~$\bf 3.43$ & $-0.79$ & $-0.63$  & $-0.34$  & ~~$0.48$ \\
$g_i\,G_i^{II}$ & $\bf -27.18$ & ~~$5.69$ & ~~$6.72$ & ~~$2.40$ & $-2.03$ \\
\hline
$3145.71 +i3.57$ & $D^* N$ & $\rho \Sigma_c$ & $\rho \Lambda_c$ & $\omega \Sigma_c$ & $\phi \Sigma_c$\\
\hline
$g_i$ & $-0.13+i0.47$~~ & ~~$\bf 3.66-i0.08$ & $-0.15-i0.10$~~  & $-0.08-i0.05$~~  & $0.11+i0.07$ \\
$g_i\,G_i^{II}$ & $-5.36-i3.42$~~ & $\bf -47.40+i0.34$~~ & $4.89-i0.81$ & $1.02+i0.67$ & $-0.67-i0.44$~~ \\
\hline
\end{tabular}
\end{table}

We have obtained three states with $I=1$,  some of them degenerate in spin, corresponding to $\Sigma_c$ states. In total there are three states with $J=1/2$ and two with $J=3/2$. In  Ref. \cite{Romanets:2012hm} one also sees three $\Sigma_c$ states with $J=1/2$ and two states with $J=3/2$. Some of them have strong couplings to particular channels, as we also found here, but the masses of the states differ somewhat.

\section{Summary of the results}

\begin{figure}[tb]
\epsfig{file=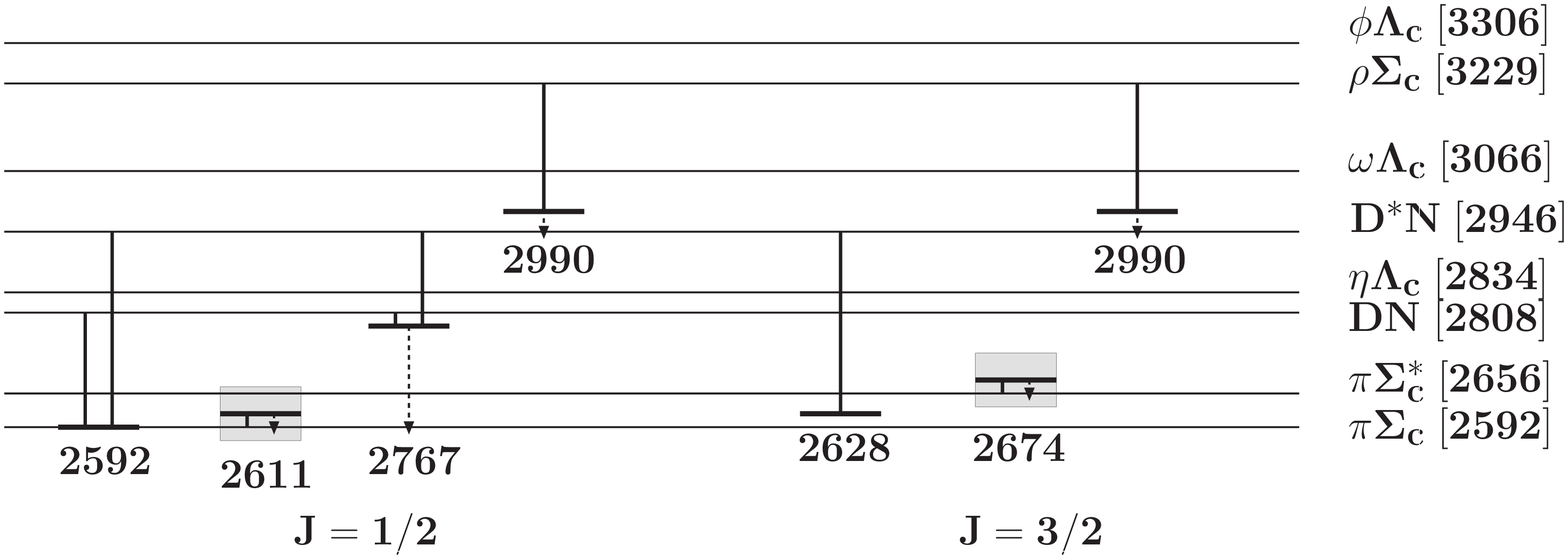, width=13cm}
\caption{Energy level diagram of the generated states with I=0. Straight and dashed vertical lines represent dominant components and main decay channels respectively. Shaded boxes stand for the width of the state. }\label{fig:I0states}
\end{figure}

\begin{figure}[tb]
\epsfig{file=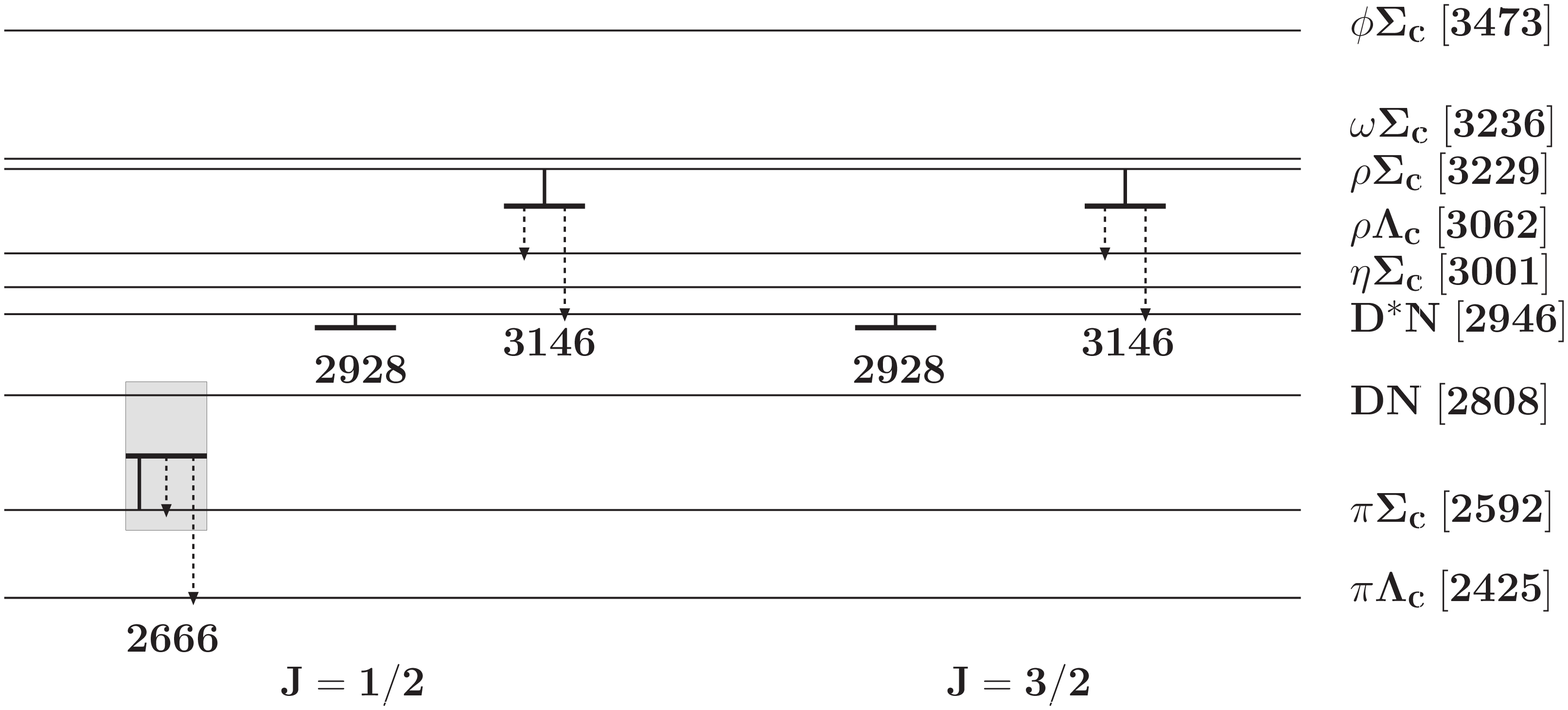, width=13cm}
\caption{Energy level diagram of the generated states with I=1. Straight and dashed vertical lines represent dominant components and main decay channels respectively. Shaded boxes stand for the width of the state. }\label{fig:I1states}
\end{figure}

Finally, we summarize here the final results that we get for the states. The results are shown in Table \ref{tab:polesum}, where we also write for reference the main channel of the state and the main decay channels. Furthermore in Figs.~\ref{fig:I0states} and \ref{fig:I1states}, the energy level diagrams of the generated states with $I=0$ and $I=1$ are depicted.

\begin{table}[H]
     \renewcommand{\arraystretch}{1.5}
     \setlength{\tabcolsep}{0.3cm}
     \caption{Energies and widths of the states obtained and the channels to which the states couple most strongly. All states have negative parity. Except for the states at $2592 \mev$ and $2628 \mev$ that have been fitted to the $\Lambda_c(2595)$ and $\Lambda_c(2625)$ respectively, and the $\pi \Sigma_c^*$ at $2674 \mev$ which is close to the $\pi \Sigma_c^*$ threshold, the other predicted energies have an estimated uncertainty of $\pm 20 \mev$.}
\label{tab:polesum}
\centering
\begin{tabular}{|c|c|c|c|c|c|}
\hline
main channel & $J$  & $I$  &   $\, (E, \, \Gamma)\ \text{[MeV]} \,$  & Exp. &  main decay channels  \\
\hline
$\frac{1}{\sqrt{2}}(D N-D^* N)$, \, $\pi \Sigma_c$  & $1/2$   & $0$ & $2592, \, 1$ &  $\Lambda_c(2595)$ & $\pi \Sigma_c$ \\
\hline
$\pi \Sigma_c$  & $ 1/2$ &$0$ & $2611, \, 106$ & -  & $\pi \Sigma_c$\\
\hline
$\frac{1}{\sqrt{2}}(D N+D^* N)$  & $1/2$ &$0$ & $2767, \, 2$ & -  &$\pi \Sigma_c$ \\
\hline
$D^* N$  &  $3/2$ &$0$  & $2628, \, 0$ &  $\Lambda_c(2625)$  & -\\
\hline
$\pi \Sigma_c^*$  &  $3/2$ &$0$  & $2674, \, 111$ & -   &$\pi \Sigma_c^*$\\
\hline
$\rho \Sigma_c$  &  $1/2, 3/2$  &$0$  & $2990, \, 2$ & $\Lambda_c (2940)?$   & $D^* N$\\
\hline
$\pi \Sigma_c$  & $1/2$ &$1$  & $2666, \, 272$ & -   &$\pi \Sigma_c, ~\pi \Lambda_c$\\
\hline
$D^* N$  &  $ 1/2, 3/2$ &$1$ & $2928, \, 0$ & -  & -\\
\hline
$\rho \Sigma_c$  & $ 1/2, 3/2$ &$1$ & $3146, \, 7$ & -  & $D^* N, ~\rho \Lambda_c$\\
\hline
\end{tabular}
\end{table}

In summary, we predict six states with $I=0$, two of them corresponding to the $\Lambda_c(2595)$ and $\Lambda_c(2625)$, and three states with $I=1$, some of them degenerate in spin. The energies of the states range from about 2592 MeV to 3146 MeV.

   It might seem at first sight that this is a large number of states, but we must recall that for the analogous  sector of baryon strange states one finds within the same range of difference of energies six $\Lambda$ states with spin and parity $J^P=1/2^-, 3/2^-, 5/2^-$ and six $\Sigma$ states with the same spin and parity, most of which could be reproduced as dynamically generated states of meson-baryon or vector-baryon \cite{revhidden,ramosoller}.

   It is interesting to see in the tables that the states found have components of $\eta (\omega, \phi)$-baryon, or $\pi \Lambda_c, \rho \Lambda_c$, but none of the states is largely made by these components, unlike some dynamically generated baryons in the light sector \cite{inoue, angelsvec}. The nondiagonal transitions from $DN, D^* N$ to such channel justify the presence of these components in the wave functions of the states found. However such transitions are suppressed because they involve the heavy $D^*$ exchange and they are penalized by the large $D^*$ square mass in the $D^*$ propagator, see Eq. \eqref{eq:approximation}. On the other hand, the $\pi$-$\Sigma_c$ and $\rho$-$\Sigma_c$ channels could develop states of $\pi$-$\Sigma_c$,  $\rho$-$\Sigma_c$ nature because of the big diagonal matrix elements of the interaction in these channels, as seen in the tables of the Appendix.

   For the moment there is only one $\Sigma_c$ state reported in Ref. \cite{pdg} around 2800 MeV. The state, however, has no spin nor parity assigned. While there are several states in Table \ref{tab:polesum} close in energy to this state, it is worth quoting that the  width of the experimental state is around 75 MeV, which is far away
from the $\Gamma =0,272$ MeV, that we find for the likely states in Table
\ref{tab:polesum} according to the mass. We would tentatively conclude that the experimental state corresponds most likely to a positive parity state.
On the other hand, the reported state $\Lambda_c (2940)$ with $\Gamma=17^{+8}_{-6}$ MeV \cite{pdg} which has no spin parity associated, could correspond to the spin degenerate $\Lambda_c$ state that we find at 2990 MeV with small width. In other approaches that use a constituent quark model \cite{entem} a $D^* N$ structure is suggested for this state. However, as shown in Tables \ref{tab:giDsNAnomI0b}, \ref{tab:giJ3/2AnomI0b}, this state, which has some coupling to $D^* N$, couples mostly to $\rho \Sigma_c$. The states dominated by $D^* N$ in our approach appear more bound.

\section{Conclusions}

In this work we studied the interaction of $D N$ and $D^* N$ states with its coupled channels using dynamics extrapolated from the light quark sector to the heavy one. The starting point was to consider the heavy quarks as spectators in the dominant terms of the interaction. The interaction was extracted mapping from the light sector and respecting the rules of heavy quark spin symmetry,
which in the local hidden gauge approach correspond to the exchange of light vectors.
Hence, an extrapolation of the results of the local hidden gauge approach was used.
The formalism is extended to allow for heavy vector ($D^*$) exchange in transitions that play a minor role in the problem. For this, the local hidden gauge approach is extrapolated  to SU(4) but only in the elementary vertices. SU(4) is then broken when actual masses are used in the mechanism of the interaction. The mixing of the $PB$ and $VB$ states was done through pion exchange. With these ingredients,
we studied the interaction of the $D N$ and $D^* N$ with their coupled channels $\pi \Sigma_c$, $\pi \Lambda_c$, $\eta \Sigma_c$ (for the $D N$);  $\rho \Sigma_c$, $\omega \Lambda_c$, $\phi \Lambda_c$, $\rho \Sigma_c^*$, $\omega \Sigma_c^*$, $\phi \Sigma_c^*$ (for the $D^* N$), and we searched for poles of the scattering matrix in different states of spin and isospin. We found six states in $I=0$, with one of them degenerate in spin $J=1/2,\ 3/2$, and three states in $I=1$,  two of them degenerate in spin $J=1/2,\ 3/2$. The coupling of the states to the different channels, together with their wave function at the origin, were evaluated to show  which is the weight of the different building blocks in those molecular states. In particular, two of the states, one with spin $1/2$ that couples mostly to a combination of $DN$ and $D^* N$, and a second one with spin $3/2$ that couples mostly to $D^* N$ were associated to the experimental ones, $\Lambda_c(2595)$ and $\Lambda_c(2625)$ respectively. The rest of states are so far predictions, with a number of states that is similar to the one of negative parity $\Lambda$ and $\Sigma$ states in the strange sector. In particular we predict another state made of $DN$, $D^* N$, orthogonal to the $\Lambda_c(2595)$, with a mass around $2767 \mev$.
We think that the use of realistic dynamics, with strict respect of heavy quark spin-flavor symmetry, renders the results obtained rather solid and they should serve as a guideline for future experiments searching for baryon states with open charm.

\section*{Acknowledgments}

This work is partly supported by the Spanish Ministerio de Economia y Competitividad and European FEDER funds under Contract No. FIS2011-28853-C02-01 and the Generalitat Valenciana in the program Prometeo, 2009/090. We acknowledge the support of the European Community-Research Infrastructure Integrating Activity Study of Strongly Interacting Matter (Hadron Physics 3, Grant No. 283286) under the Seventh Framework Programme of the European Union.
This work is also partly supported by the National Natural Science Foundation of China under Grant No. 11165005.

\section*{Appendix: Tables for the $C_{ij}$ coefficients}
\begin{appendix}
\setcounter{table}{0}
\renewcommand{\thetable}{A\arabic{table}}

\begin{table}[H]
     \renewcommand{\arraystretch}{1.5}
     \setlength{\tabcolsep}{0.4cm}
     \caption{$C_{ij}$ coefficients for $D N$ and coupled channels with $I=0$ and $J^P=1/2^-$.}
\label{tab:vij0}
\centering
\begin{tabular}{c|ccc}
$C_{ij}$ & $D N$ &  $\pi \Sigma_c$ & $\eta \Lambda_c$  \\
\hline
$D N$ &  3 & $-\sqrt{\frac{3}{2}}$ & $\frac{3}{\sqrt{2}}$   \\
$\pi \Sigma_c$ &   & 4  & 0   \\
$\eta \Lambda_c$ &  &  & 0
\end{tabular}
\end{table}

\begin{table}[H]
     \renewcommand{\arraystretch}{1.5}
     \setlength{\tabcolsep}{0.4cm}
     \caption{$C_{ij}$ coefficients for $D N$ and coupled channels with $I=1$ and $J^P=1/2^-$.}
\label{tab:vij1}
\centering
\begin{tabular}{c|cccc}
$C_{ij}$ & $D N$ &  $\pi \Sigma_c$ &  $\pi \Lambda_c$ & $\eta \Sigma_c$  \\
\hline
$D N$ &  1 & $-1$ & $-\sqrt{\frac{3}{2}}$ & $-\sqrt{\frac{3}{2}}$   \\
$\pi \Sigma_c$ &   & 2  & 0  & 0 \\
$\pi \Lambda_c$ &  &  & 0 & 0  \\
$\eta \Sigma_c$&  &  & & 0
\end{tabular}
\end{table}

\begin{table}[H]
     \renewcommand{\arraystretch}{1.5}
     \setlength{\tabcolsep}{0.4cm}
     \caption{$C_{ij}$ coefficients for $D^* N$ and coupled channels with $I=0$, and $J^P = 1/2^-, \, 3/2^-$.}
\label{tab:vij0j13}
\centering
\begin{tabular}{c|cccc}
$C_{ij}$ & $D^* N$ &  $\rho \Sigma_c$ &  $\omega \Lambda_c$ & $\phi \Lambda_c$  \\
\hline
$D^* N$ &  3 & $-\sqrt{\frac{3}{2}}$ & $\sqrt{\frac{3}{2}}$ & $-\sqrt{3}$   \\
$\rho \Sigma_c$ &   & 4  & 0  & 0 \\
$\omega \Lambda_c$ &  &  & 0 & 0  \\
$\phi \Lambda_c$&  &  & & 0
\end{tabular}
\end{table}

\begin{table}[H]
     \renewcommand{\arraystretch}{1.5}
     \setlength{\tabcolsep}{0.4cm}
     \caption{$C_{ij}$ coefficients for $D^* N$ and coupled channels with $I=1$, and $J^P = 1/2^-, \, 3/2^-$.}
\label{tab:vij1j13}
\centering
\begin{tabular}{c|ccccc}
$C_{ij}$ & $D^* N$ &  $\rho \Sigma_c$ &  $\rho \Lambda_c$ & $\omega \Sigma_c$   & $\phi \Sigma_c$ \\
\hline
$D^* N$ &  1 & $-1$ & $-\sqrt{\frac{3}{2}}$ & $-\sqrt{\frac{1}{2}}$ & 1   \\
$\rho \Sigma_c$ &   & 2  & 0  & 0  & 0  \\
$\rho \Lambda_c$ &  &  & 0 & 0  & 0   \\
$\omega \Sigma_c$ &  &  &  & 0  & 0    \\
$\phi \Sigma_c$ &  &  &  &  & 0
\end{tabular}
\end{table}

\begin{table}[H]
     \renewcommand{\arraystretch}{1.5}
     \setlength{\tabcolsep}{0.4cm}
     \caption{$C_{ij}$ coefficient for $\pi \Sigma_c^*$ with $I=0$ and $J^P=3/2^-$.}
\label{tab:vij0j3}
\centering
\begin{tabular}{c|c}
$C_{ij}$ &  $\pi \Sigma_c^*$   \\
\hline
$\pi \Sigma_c^*$ &  4
\end{tabular}
\end{table}

The zeros in the diagonal $\eta \Lambda_c, \eta \Sigma_c, \omega \Lambda_c, \phi \Lambda_c$ are easy to explain since the $\rho^0, \omega, \phi$ exchange violate $C$-parity. For $\pi \Lambda_c$ the $\rho \Lambda_c \Lambda_c$ is forbidden by isospin and $\omega, \phi$ do not couple to pions.

\end{appendix}

\end{document}